\documentclass[aps,prbtwocolumn,superscriptaddress,citeautoscript,amssymb,reprint,showpacs,floatfix,longbibliography]{revtex4-2}
\usepackage[ansinew]{inputenc}
\usepackage[T1]{fontenc}
\usepackage{float}
\usepackage{amsmath}
\usepackage[normalem]{ulem}
\usepackage{mathptmx}
\usepackage{mathrsfs}
\usepackage[english]{babel}
\usepackage{graphicx}
\usepackage[colorlinks=true,linkcolor=blue,citecolor=blue,urlcolor=blue]{hyperref}
\usepackage[usenames,dvipsnames]{color}
\usepackage{natbib}
\usepackage{etoolbox}

\usepackage{makecell}
\usepackage[thinspace,thinqspace]{SIunits}

\newcommand{\CRA}{CeRh$_2$As$_2$}
\newcommand{\Tc}{$T_{\textrm{c}}$}
\newcommand{\To}{$T_{\textrm{0}}$}
\newcommand{\muo}{$\mu_{\textrm{0}}$}

\makeatletter
\patchcmd{\@bibitem}{\@biblabel}{\@biblabel}{}{}
\patchcmd{\@bibitem}{``}{\ignorespaces}{}{}
\patchcmd{\@bibitem}{''}{\ignorespaces}{}{}
\makeatother

\begin{document}

\author{M.~Pfeiffer}
\thanks{These authors contributed equally.}
\affiliation{Institute for Solid State and Materials Physics, TU Dresden University of Technology, 01062 Dresden, Germany}
\affiliation{Max Planck Institute for Chemical Physics of Solids, 01187 Dresden, Germany}

\author{K.~Semeniuk}
\thanks{These authors contributed equally.}
\affiliation{Max Planck Institute for Chemical Physics of Solids, 01187 Dresden, Germany}

\author{J.~F.~Landaeta}
\affiliation{Institute for Solid State and Materials Physics, TU Dresden University of Technology, 01062 Dresden, Germany}
\affiliation{Max Planck Institute for Chemical Physics of Solids, 01187 Dresden, Germany}

\author{R.~Borth}
\affiliation{Max Planck Institute for Chemical Physics of Solids, 01187 Dresden, Germany}

\author{C.~Geibel}
\affiliation{Max Planck Institute for Chemical Physics of Solids, 01187 Dresden, Germany}

\author{M.~Nicklas}
\affiliation{Max Planck Institute for Chemical Physics of Solids, 01187 Dresden, Germany}

\author{M.~Brando}
\affiliation{Max Planck Institute for Chemical Physics of Solids, 01187 Dresden, Germany}

\author{S.~Khim}
\affiliation{Max Planck Institute for Chemical Physics of Solids, 01187 Dresden, Germany}

\author{E.~Hassinger}
\thanks{correspondence should be addressed to konstantin.semeniuk@cpfs.mpg.de and elena.hassinger@tu-dresden.de}
\affiliation{Institute for Solid State and Materials Physics, TU Dresden University of Technology, 01062 Dresden, Germany}
\affiliation{Max Planck Institute for Chemical Physics of Solids, 01187 Dresden, Germany}

\title{Pressure-tuned Quantum Criticality in the Locally Noncentrosymmetric Superconductor \CRA}
\date{\today}

\begin{abstract}

The unconventional superconductor \CRA\ (critical temperature \Tc$\approx0.4$\,K) displays an exceptionally rare magnetic-field-induced transition between two distinct superconducting (SC) phases, proposed to be states of even and odd parity of the SC order parameter, which are enabled by a locally noncentrosymmetric structure. The superconductivity is preceded by a phase transition of unknown origin at $T_{0}=0.5$\,K. Electronic low-temperature properties of \CRA\ show pronounced non-Fermi-liquid behavior, indicative of a proximity to a quantum critical point (QCP). The role of quantum fluctuations and normal state orders for the superconductivity in a system with staggered Rashba interaction is currently an open question, pertinent to explaining the occurrence of the two-phase superconductivity. In this work, using measurements of resistivity and specific heat under hydrostatic pressure, we show that the \To\ order vanishes completely at a modest pressure of $P_0\approx0.5$\,GPa, revealing a QCP. In line with the quantum criticality picture, the linear temperature dependence of the resistivity at $P_{0}$ evolves into a Fermi-liquid quadratic dependence as quantum critical fluctuations are suppressed by increasing pressure. Furthermore, the domelike behavior of \Tc\ around $P_{0}$ implies that the fluctuations of the \To\ order are involved in the SC pairing mechanism.

\end{abstract}
\maketitle

\begin{figure}[t]
	\includegraphics[width=\columnwidth]{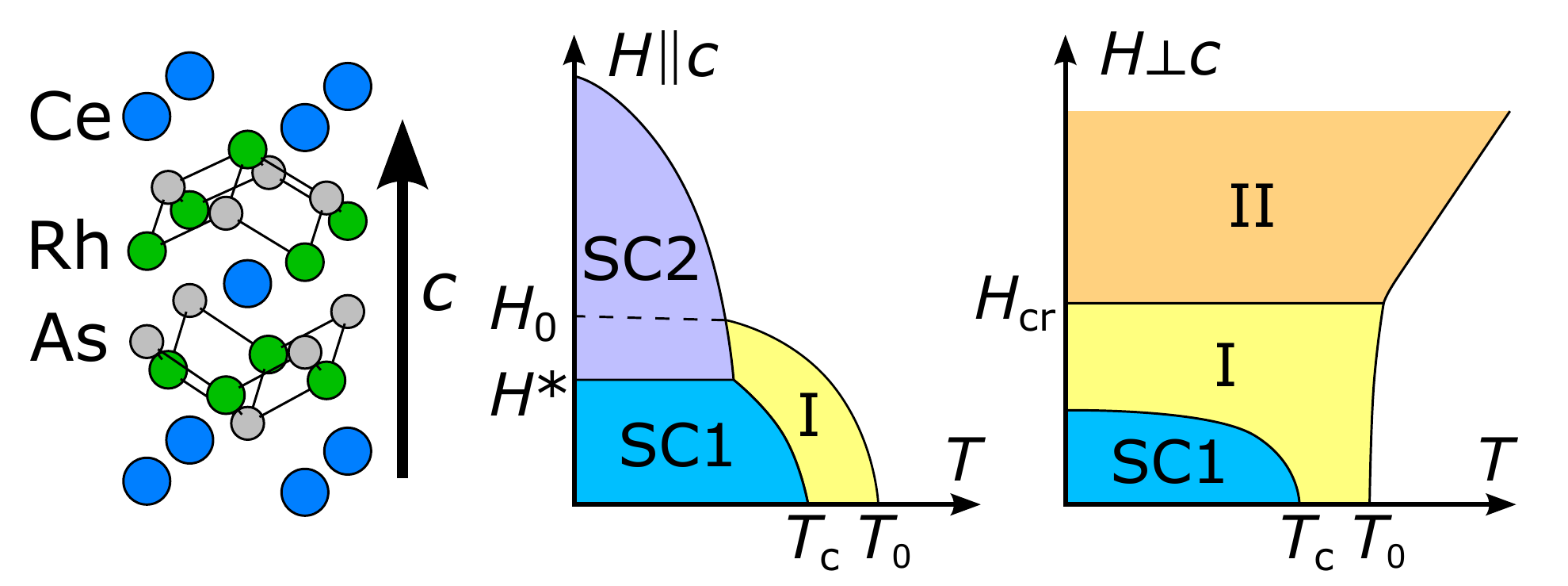}
	\caption{Crystalline structure and low-temperature states of \CRA\ for magnetic field parallel and perpendicular to the $c$ axis. SC1/SC2 denotes the low-field or high-field superconducting state, I/II is the nonsuperconducting state phase~I or phase~II. The dashed line marks a so-far undetected segment of the phase boundary of phase~I, terminating at a field $H_{0}$.}
\label{fig:intro}
\end{figure}

\textit{Introduction}---Strong electron-electron interactions in clean crystalline systems often lead to phenomena that challenge our understanding of condensed matter. These typically emerge at thresholds of second order phase transitions, particularly near quantum critical points (QCPs), which are an established source of unconventional superconductivity and non-Fermi-liquid behavior. The recently discovered multiphase superconductor \CRA~\cite{khim2021} displays strong electronic correlations and non-Fermi liquid properties but an associated QCP has not been identified so far. \CRA\ stands out among superconducting (SC) Kondo-lattice systems by crystallizing in a nonsymmorphic CaBe$_2$Ge$_2$-type structure (space group \#129, $P4/nmm$, see Fig.~\ref{fig:intro})~\cite{madar1987}. The inversion symmetry is absent for Ce sites, but is preserved globally, causing staggered Rashba spin-orbit interaction, which is believed to lead to the multi-phase superconductivity~\cite{yoshida2012,khim2021}. However, it is unclear, to what extent the established paradigm of unconventional superconductivity emerging at a QCP is upheld in a material such as \CRA. Here, we address these questions by tuning the electronic correlations in the system with hydrostatic pressure.

Magnetic and chanrge transport properties of \CRA\ show typical Kondo lattice behavior (Kondo temperature ${T_{\mathrm{K}}\approx30}$\,K)~\cite{khim2021}. At low temperature ($T$) and ambient pressure ($P$), the non-Fermi-liquid behavior clearly manifests in the specific heat ($C$), with $C/T\sim T^{-0.6}$ above the ordering temperature, also observed for quantum critical Yb systems~\cite{steppke2013,custers2003}. The large Sommerfeld coefficient of ${\sim1}$\,J\,mol$^{-1}$\,K$^{-2}$ at 0.5\,K implies colossal effective charge carrier masses. Furthermore, the low-temperature resistivity ($\rho$) has a sub-linear temperature dependence, approximated by ${\rho-\rho_0\propto\sqrt{T}}$ near 1\,K ($\rho_0$~-- residual resistivity).

The low-temperature states of \CRA\ are indicated in schematic phase diagrams in Fig.~\ref{fig:intro}. The material is a superconductor with the critical temperature $T_{\mathrm{c}}\approx0.4$\,K and hosts two distinct SC phases: the low-field or high-field phase SC1/SC2. The SC1-SC2 transition is induced by the magnetic field of $\mu_{0}H^{*}=4$\,T, applied along the $c$ axis of the tetragonal lattice (the out-of-plane direction). The origin of the phenomenon is yet to be revealed by a direct experiment. The dominant hypothesis views SC1 and SC2 as phases of even and odd parity of the SC order parameter, respectively~\cite{khim2021,landaeta2022}. Alternatively, another ordered state may coexist with the superconductivity at fields below $H^{*}$, but not above~\cite{ogata2023,machida2022}.

\begin{figure}
    \includegraphics[width=\columnwidth]{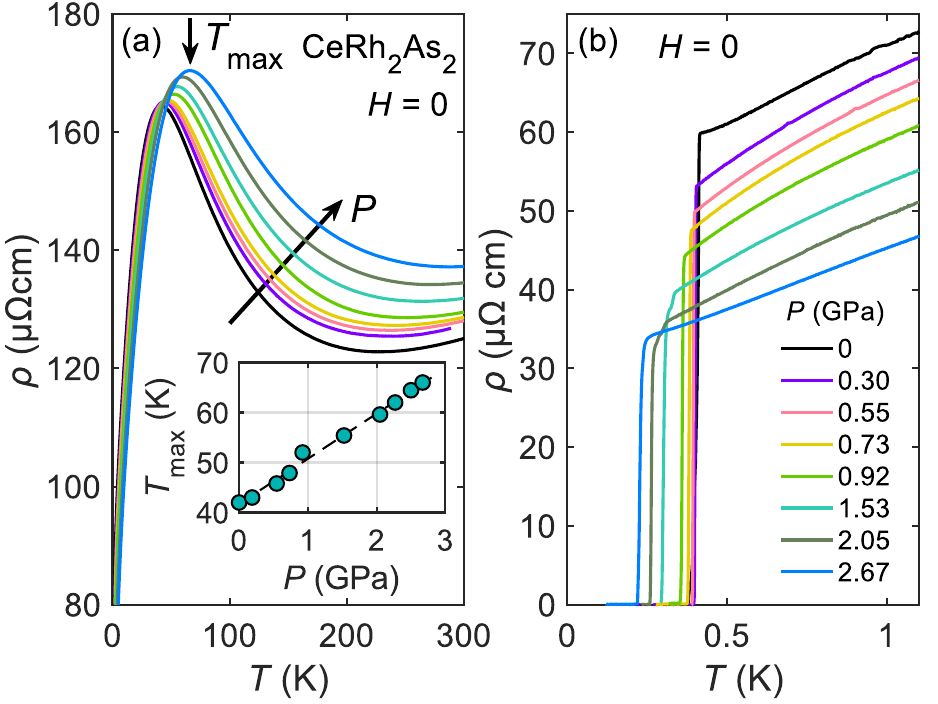}
    \caption{Effect of pressure ($P$) on the temperature dependence of resistivity ($\rho(T)$) of \CRA. (a) Resistivity in a broad temperature range. Pressures are given in the panel (b). Inset: the Kondo coherence peak temperature $T_{\mathrm{max}}$ against pressure. (b) Low-temperature resistivity at different pressures.}
    \label{fig:ZeroFieldData}
\end{figure}

At the temperature $T_{0}\approx0.5$\,K, \CRA\ undergoes a second order transition into a state ``Phase~I,'' of currently unknown microscopic origin~\cite{hafner2022,mishra2022,semeniuk2023}. The anomaly manifests as a resistivity upturn upon cooling, which varies in magnitude with the current flow direction~\cite{mishra2022}, suggesting a partial gapping of the Fermi surface (FS). Phase~I has so far remained invisible in magnetic susceptibility and magnetization measurements, and a quadrupolar nature of it has been considered~\cite{hafner2022} (the crystal-electric-field ground state at the Ce ion is a quasi-quartet with quadrupolar degrees of freedom~\cite{christovam2024}). The ordered states of \CRA\ display a highly anisotropic response to magnetic field (Fig.~\ref{fig:intro}): for a $c$-axis field $H_{\parallel}$, \To\ is suppressed at approximately 6\,T; for an $ab$-plane field $H_{\perp}$ (the in-plane direction) the SC2 phase is absent and \To\ is stable, but at $\mu_{0}H_{\mathrm{cr}}\approx9$\,T the system undergoes a transition from phase~I to a state ``Phase~II,'' the critical temperature of which steadily increases with field up to at least 18\,T~\cite{hafner2022}.

Studies of $^{75}$As nuclear magnetic and quadrupolar resonance (NMR and NQR) in \CRA\ have shown the presence of internal fields within the SC state which imply an antiferromagnetic (AFM) order with the critical temperature $T_{\mathrm{N}}$ of at least 0.25\,K~\cite{kitagawa2022,kibune2022,ogata2023}. This transition has not been detected by other probes yet.

While being a unique multiphase superconductor, \CRA\ also exhibits signatures typical of many Kondo-lattice heavy-fermion superconductors, where the SC phase surrounds a magnetic QCP~\cite{mathur1998}. However, earlier thermal expansion and heat capacity data suggested an initial increase of \To\ with pressure and hence a departure from the putative QCP of phase~I~\cite{hafner2022}, contrasting the standard scenario for AFM Ce-based heavy-fermion systems~\cite{knebel2011}. In this study, we subjected \CRA\ to hydrostatic pressure and investigated its effects on the stability of the ordered states and the strongly correlated behavior via measurements of heat capacity and electrical resistivity along the $ab$ plane (the methods used are described in Supplemental Material [SM]).

In contrast to the predictions, we found that \To\ decreases with pressure and vanishes at approximately 0.5\,GPa, revealing a QCP. This suppression is accompanied by a dome-like evolution of \Tc\ around the critical pressure and a gradual recovery of a $\sim T^{2}$ dependence of the low-temperature resistivity away from the QCP, in accordance with the usual scenario. We note similarities and differences between \CRA\ and other Ce-based systems in this context.

\textit{Results}---In Fig.~\ref{fig:ZeroFieldData}(a) we show resistivity of \CRA\ against temperature up to 300\,K at different pressures. The presented data were measured on a sample from the same batch of \CRA\ single crystals as the one used in the initial studies of the material~\cite{khim2021,landaeta2022,hafner2022,kitagawa2022,kibune2022,ogata2023} (see SM for notes on the batches of \CRA\ used in this work). The Kondo-coherence peak shifts linearly from 42\,K at ambient pressure to 66\,K at 2.67\,GPa (8.9 K/GPa). The enhancement of the coherence-peak temperature (and hence of $T_{\mathrm{K}}$) with pressure is common for Ce-based Kondo-lattice systems~\cite{goltsev2005}, implying that the paramagnetic ground state becomes favored~\cite{parks1977}. Figure~\ref{fig:ZeroFieldData}(b) shows the data in the low-temperature interval (${T<1}$\,K). In this region, the normal-state resistivity is strongly reduced under pressure: at 0.5\,K, $\rho$ decreases from $\sim$60 to $\sim$35\,$\micro\Omega$\,cm between 0 and 2.67\,GPa. For a given temperature, the curvature of $\rho(T)$ in the normal state increases with pressure, which will be addressed quantitatively later.

\begin{figure}[!t]
     \includegraphics[width=\columnwidth]{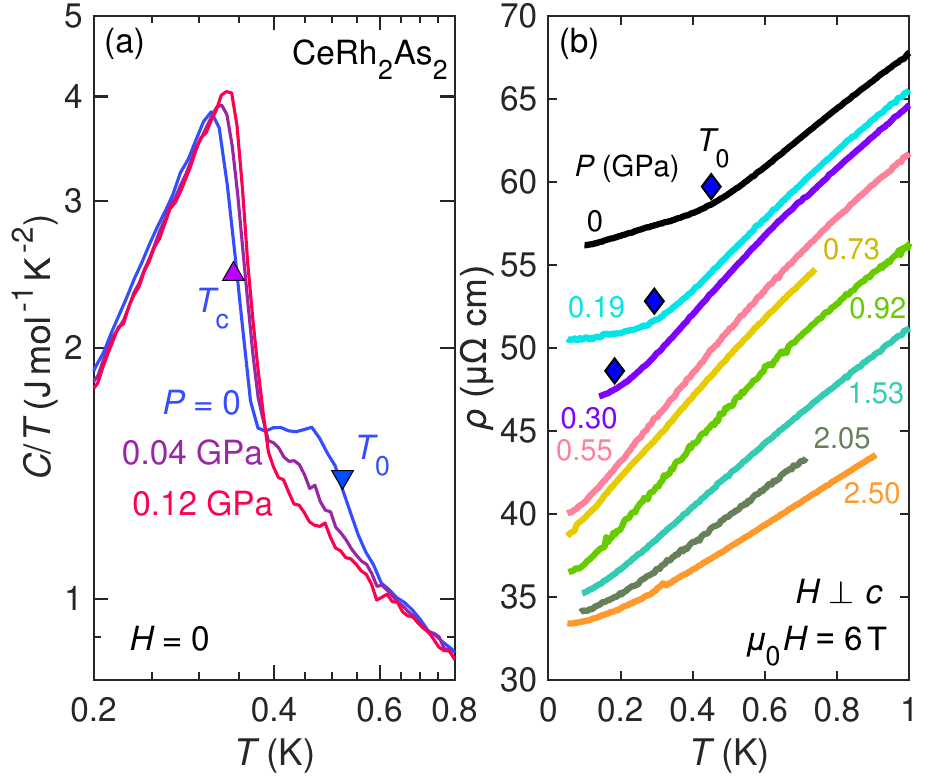}
     \caption{Phase~I of \CRA\ under pressure. (a) Temperature ($T$) dependence of zero-field heat capacity $C$ divided by $T$ for a high-quality batch of \CRA. (b) Resistivity $\rho$ against $T$ at different pressures in a 6\,T in-plane magnetic field $H$. The temperature \To\ is defined as the point of maximum curvature of $\rho(T)$ (blue diamonds).}
     \label{fig:T0_data}
\end{figure}

We now examine the pressure induced changes to \Tc\ and \To. At ambient pressure and in zero field, a slight upturn of $\rho(T)$ near 0.5\,K marks the resistive signature of \To\ [see Fig.~\ref{fig:ZeroFieldData}(b)]. While \Tc\ barely changes at low pressures, at merely 0.3\,GPa the upturn disappears, indicating a suppression of \To. We examine this low-pressure behavior via zero-field heat capacity measurements, conducted on crystals of improved quality. Data in Fig.~\ref{fig:T0_data}(a) show clear anomalies at \Tc\ and \To\ at ambient pressure. It is known from the previous measurements~\cite{semeniuk2023}, that the smaller specific heat anomaly and the upturn in resistivity correspond to the same \To\ transition---both occur at the same temperature within a given batch of \CRA\ and show identical magnetic-field dependence. The response to pressure is also consistent: the \To\ anomaly, initially located at 0.53\,K, shifts down in temperature at 0.04\,GPa and becomes obscured by the larger \Tc\ anomaly at 0.12 GPa, indicating $dT_{0}/dP\approx-1$\,K/GPa. The behavior of the \To\ transition inside the SC state is unclear at the moment, and can differ significantly depending on the nature of the interaction between the superconductivity and phase~I~\cite{semeniuk2023}.

\begin{figure}
	\includegraphics[width=\columnwidth]{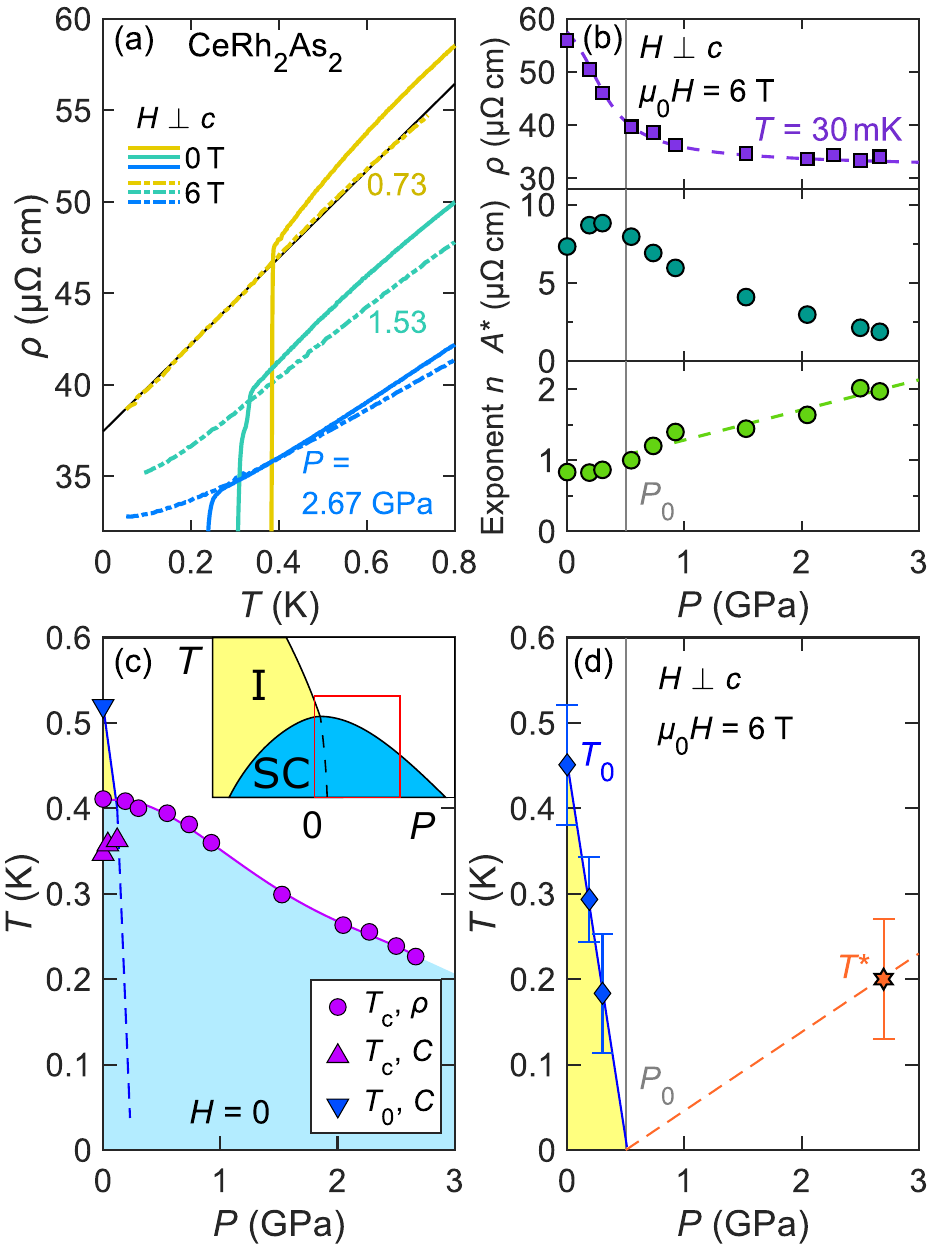}
    \caption{Ordered states and electronic correlations in \CRA\ under pressure ($P$). (a) Resistivity $\rho$ against temperature $T$ at selected pressures. (b) Top: pressure dependence of $\rho$ at 30\,mK. $A^*$ parameter (middle) and exponent $n$ (bottom) against pressure, obtained by fitting the data at a 6\,T in-plane field with a function $\rho(T)=\rho_{0}+A^*(T/T_{\mathrm{ref}})^{n}$ ($T_{\mathrm{ref}}=0.3$\,K, $\rho_{0}$---residual resistivity; details provided in the text and Supplemental Material). (c) Zero-field phase diagram. The triangular markers correspond to the heat capacity ($C$) data. The inset shows a possible schematic phase diagram for a broader tuning parameter range. The red outline marks the pressure-temperature range explored in this work for \CRA. (d) Phase diagram for a 6\,T in-plane field. The system crosses over into the Fermi-liquid regime below $T^*$.}
\label{fig:normal_state_summary}
\end{figure}

In the phase diagrams shown in Fig.~\ref{fig:intro}, one can see that \To\ depends rather weakly on $H_{\perp}$ before phase~II is induced (we reproduced the known $T$-$H_{\perp}$ phase diagram~\cite{hafner2022,mishra2022} in the current study, see SM). Hence, it is possible to trace \To\ at higher pressures by applying an in-plane field, such that the superconductivity is suppressed. In Fig.~\ref{fig:T0_data}(b), we show $\rho(T)$ at $\mu_{0}H_{\perp}=6$\,T. One can indeed identify \To\ by a change of slope of $\rho(T)$. Applying pressure shifts \To\ down in temperature at a rate of -1\,K/GPa; at 0.55 GPa we only observe the onset of the upturn below 0.1\,K, while no signs of \To\ are visible at 0.73 GPa. Such a reduction of \To\ agrees with the heat capacity data and likely reflects the bulk behavior at zero field, when \To\ is higher than \Tc. An extrapolation of $T_{0}(P)$ past 0.3\,GPa at \muo$H_{\perp}=6$\,T suggests that phase~I should vanish completely at a critical pressure $P_{0}\approx0.5$\,GPa [Fig.~\ref{fig:normal_state_summary}(d)]. The same analysis performed for \muo$H_{\perp}=4$\,T results in a slightly smaller $P_{0}$ of $0.41$\,GPa (see SM), consistent with weakly varying $T_{0}(H_{\perp})$ at these fields. In higher-quality \CRA\ crystals, \To\ can be resolved more clearly in $\rho(T)$ even at zero field~\cite{semeniuk2023}. A limited set of zero-field measurements done on these crystals also points at \To\ decreasing with pressure (see SM).

Concerning the superconductivity, \Tc\ is initially weakly enhanced by pressure, as seen in the heat capacity data [Fig~\ref{fig:T0_data}(a)]. The resistive \Tc, on the other hand, is nearly constant below 0.55\,GPa and decreases steadily at higher pressures. Different values of \Tc\ in specific heat and resistivity have been observed before \cite{khim2021,semeniuk2023}, and this mismatch is probably caused by sample inhomogeneity, as discussed in SM. We expect that a maximum in $T_{\mathrm{c}}(P)$ occurs near the crossing of \Tc\ and \To, resembling the interplay of SC and magnetic critical temperatures in other Ce-based systems~\cite{knebel2011}. This can explain why the initial increase of \Tc\ is not apparent in the older-batch resistivity data where \To\ is lower and the peak of the $T_{\mathrm{c}}(P)$ dome is near $P=0$. Resistivity measurements on crystals of improved quality show a positive $dT_{\mathrm{c}}/dP$ value at $P=0$ (see SM). Across the entire studied pressure range of 2.7\,GPa , \Tc\ shows an overall decrease from 0.41 to 0.23\,K. An independent high pressure study of \CRA~\cite{siddiquee2023} reported a discontinuity in $T_{\mathrm{c}}(P)$ at 2.5\,GPa, but the authors suspected it to be caused by a sudden loss of hydrostaticity due to the solidification of the pressure transmitting medium during pressure increase. This suspicion seems justified, given that we avoid the room-temperature solidification with our chosen medium (polyethylsiloxane) and observe a continuous change in \Tc\ in the relevant pressure range.

When it comes to the unordered state, the most noteworthy features of the pressure dependence of $\rho(T)$ are the strong reduction of the low-temperature resistivity, as well as the recovery of its positive curvature. Shown in Fig.~\ref{fig:normal_state_summary}(a) for a selected set of pressures, these features are visible above \Tc\ and become particularly clear when the normal-state resistivity is extended to the lowest achieved temperature under applied magnetic field. The reduction of resistivity extends all the way to the residual value [$\rho$ at 30\,mK and 6\,T is plotted in the top panel of Fig.~\ref{fig:normal_state_summary}(b)] and likely involves two contributions. First, the change could be caused by the diminished scattering of charge carriers due to a weakening of quantum fluctuations near a QCP. Scattering of this origin should be the strongest at the critical pressure, implying a peak in $\rho_0$ at $P_0$. However, in our case, the residual resistivity is clearly affected by the existence of phase~I at $P<P_{0}$. This leads to the second cause: the decrease of $\rho$ can also be attributed to a possible increase of the charge carrier concentration due to pressure-induced changes to the FS. These may include the closing of a putative partial FS gap associated with phase~I, but not as the sole reason, since the trend is also apparent above \To.

At 0.73\,GPa (close to $P_0$), with the 6\,T in-plane magnetic field applied, $\rho(T)$ demonstrates a $T$-linear dependence below 0.6\,K and down to the lowest temperature [see Fig.~\ref{fig:normal_state_summary}(a)]. For higher pressures, a positive curvature of $\rho(T)$ develops at low $T$. We examined this trend quantitatively by fitting a power-law dependence to the low-temperature resistivity. We note the following: (i) the shape of $\rho(T)$ cannot be well approximated by a fixed-exponent power law when $T$ covers a large interval of the order of 1\,K; (ii) the expression $\rho(T)=\rho_{0}+AT^n$, typically used for Fermi-liquid metals ($n=2$) becomes inappropriate when $n$ changes with pressure. To deal with the first issue, we fitted the model to a narrow temperature interval of 250\,mK, extending to the lowest possible temperature in the unordered state. Regarding the second point, we used a modified power-law expression $\rho(T)=\rho_{0}+A^*(T/T_{\mathrm{ref}})^n$, with $T_{\mathrm{ref}}=0.3$\,K (the upper bound of the fitting interval). For such a form, the parameter $A^*$ corresponds to an increment in $\rho$ between $T=0$ and $T_{\mathrm{ref}}$, irrespective of the value of $n$. Hence, $A^*$ acts as a qualitative measure of the strength of correlations, similarly to the conventional $A$ coefficient. While the value of $T_{\mathrm{ref}}$ is arbitrary, we set it to 0.3\,K, which is close to the upper limit of the fitting interval for $P\geq0.73$\,GPa. The resultant changes in $A^*$ and $n$ with pressure for $\mu_{0}H_{\perp}=6$\,T are shown in Fig.~\ref{fig:normal_state_summary}(b). Below 0.73\,GPa, the fitting interval is shifted to higher temperatures, above the onset of phase~I. The values of $n$ and $A^*$ for $P<0.73$\,GPa should then be treated as lower and upper bounds, respectively, when compared to the $P\geq0.73$\,GPa values ($n$ decreases with temperature, while $A^*$ increases; see SM for more details on the power-law analysis). We find that the exponent $n$ increases significantly with pressure [Fig.~\ref{fig:normal_state_summary}(b), bottom panel], being approximately 1 at 0.73\,GPa and approaching the Fermi-liquid value of 2 at 2.67\,GPa. Simultaneously, $A^*$ decreases by nearly 70\% upon increasing the pressure from $P_0$ to 2.67\,GPa [Fig.~\ref{fig:normal_state_summary}(b), middle panel]. We estimate that at 2.67\,GPa, the $\sim T^2$ dependence occurs below a crossover temperature $T^*=(0.2\pm0.07)$\,K [Fig.~\ref{fig:normal_state_summary}(d)]. The observed trends, also upheld in a higher temperature interval at zero field (see SM), indicate a development of the Fermi-liquid state and weakening of electronic correlations with pressure, both of which signify a departure from a QCP.

\textit{Discussion}---Our work shows strong evidence for a phase~I QCP at $P_{0}$. In Ce-based Kondo-lattice superconductors, typically, a weak AFM phase is suppressed by pressure and both a SC dome and non-Fermi-liquid behavior are observed near the QCP~\cite{settai2007,knebel2011}. Our results show that this general principle is also upheld in \CRA, but two distinguishing aspects have to be pointed out. First, compared to other Ce based compounds, the SC phase in \CRA\ extends over a substantially wider pressure range, particularly given its low \Tc~\cite{settai2007,knebel2011,landaeta2018}. Since $T^*\approx0.2$\,K even 2\,GPa away from the QCP, the recovery of the Fermi-liquid behavior also proceeds at a comparatively slow rate. Second, in \CRA, phase~I takes the place of the AFM state in other materials, although the origin of the \To\ transition is currently an open question.

Our measurements with two different probes directly show that $dT_{0}/dP<0$ and $dT_{\mathrm{c}}/dP>0$ at $P=0$. As mentioned in the introduction, the earlier thermal expansion study~\cite{hafner2022} predicted the critical temperatures to shift in the opposite directions with pressure, according to the Ehrenfest relation for a second order phase transition. Naturally, the laws of thermodynamics cannot be violated, but the root of this contradiction is yet to be identified. Revised measurements of thermal expansion is the next step towards resolving the issue.

The predicted positive pressure dependence of \To\ has been used as an argument against phase~I being a conventional antiferromagnetic state and in favor of a quadrupole-density wave~\cite{hafner2022}. However, the negative gradient of $T_{0}(P)$ observed here does not rule out the possibility of phase~I being a nonmagnetic density-wave state. The decrease of \To\ with pressure can then be explained by increasing $T_{\mathrm{K}}$, which leads to a wider bandwidth, smaller Lindhard response function, and, consequently, weaker FS nesting. A proper quantitative analysis of this scenario would also have to take into account changes to the crystal electric fields.

The currently available experimental findings concerning the ordered states of \CRA\ present a conundrum. Across various measurements with bulk thermodynamic probes at zero field, only two anomalies have so far remained reliably reproducible~\cite{khim2021,hafner2022,semeniuk2023,chajewski2024}: the SC transition at \Tc\ and the phase~I transition at \To. Puzzlingly, the NMR and NQR studies suggest that an AFM order sets in at a critical temperature $T_\mathrm{N}$ slightly below \Tc~\cite{kitagawa2022,kibune2022,ogata2023}. A recent heat capacity study~\cite{chajewski2024} reports an additional first-order transition, occurring just below \Tc\ for $\mu_{0}H_{\parallel}\geq3$\,T, which, however, has not been detected in any of our measurements to date. The absence of the third thermodynamic phase transition could be explained by an accidental degeneracy of \Tc\ and $T_\mathrm{N}$, but given that no splitting of the anomalies has been observed in magnetic fields or under pressure, such a scenario seems very unlikely. A possibility of $T_{\mathrm{c}}=T_{\mathrm{N}}$ being enforced by a particular coupling between the respective order parameters has been considered for \CRA~\cite{szabo2023preprint}, but was found to require the superconductivity and magnetism to have specific microscopic properties, yet to be uncovered experimentally. Another potential scenario, where the AFM order actually sets in at \To, on one hand, leaves the question of why the NMR/NQR experiments find a lower ordering temperature, but on the other hand, explains the similarity of the pressure-temperature phase diagram of \CRA\ to those of other Ce-based heavy-fermion systems and in particular CeCu$_2$Si$_2$~\cite{settai2007,knebel2011,lengyel2011,smidman2023}.

\textit{Conclusions}---Summarizing the present work on \CRA\, we observe a strong sensitivity of phase~I to pressure, with $dT_{0}/dP\approx-1$\,K/GPa, which results in the \To\ transition vanishing at $0.5$\,GPa. This pressure marks a QCP, and the associated quantum critical fluctuations drive the formation of the SC state, in accordance with the established paradigm of unconventional superconductivity. The existence of the QCP is further supported by the pressure dependence of the field-temperature phase diagrams of the superconductivity, examined in a related work~\cite{semeniuk2024}. Our data also show that hydrostatic pressure can control the electronic correlations and non-Fermi-liquid behavior in the system, but the SC state and therefore the pairing-mediating fluctuations remain rather robust even at 2.7\,GPa, well past the critical pressure of \To. The microscopic origin of phase~I remains an outstanding question.

\textit{Acknowledgments}---We are thankful to Y. Yanase, G. Knebel, D. Aoki, A. Ramires, G. Zwicknagl, C. Timm, J. Link, K. Ishida, S. Kitagawa, M. Grosche, and M.-A. M\'easson for stimulating exchange. We acknowledge funding from  Deutsche Forschungsgemeinschaft (DFG) via the CRC 1143 - Project No. 247310070 (project C10) and the Wuerzburg-Dresden cluster of excellence EXC 2147 ct.qmat Complexity and Topology in Quantum Matter - Project No. 390858490.


\begin{center}
\textbf{\large Supplemental Material}
\end{center}

\setcounter{equation}{0}
\setcounter{figure}{0}
\setcounter{table}{0}
\makeatletter
\renewcommand{\theequation}{S\arabic{equation}}
\renewcommand{\thefigure}{S\arabic{figure}}
\renewcommand{\thetable}{S\arabic{table}}

\section{Methods}

\subsection{Sample preparation}

\begin{figure}[t]
        \includegraphics[width=\linewidth]{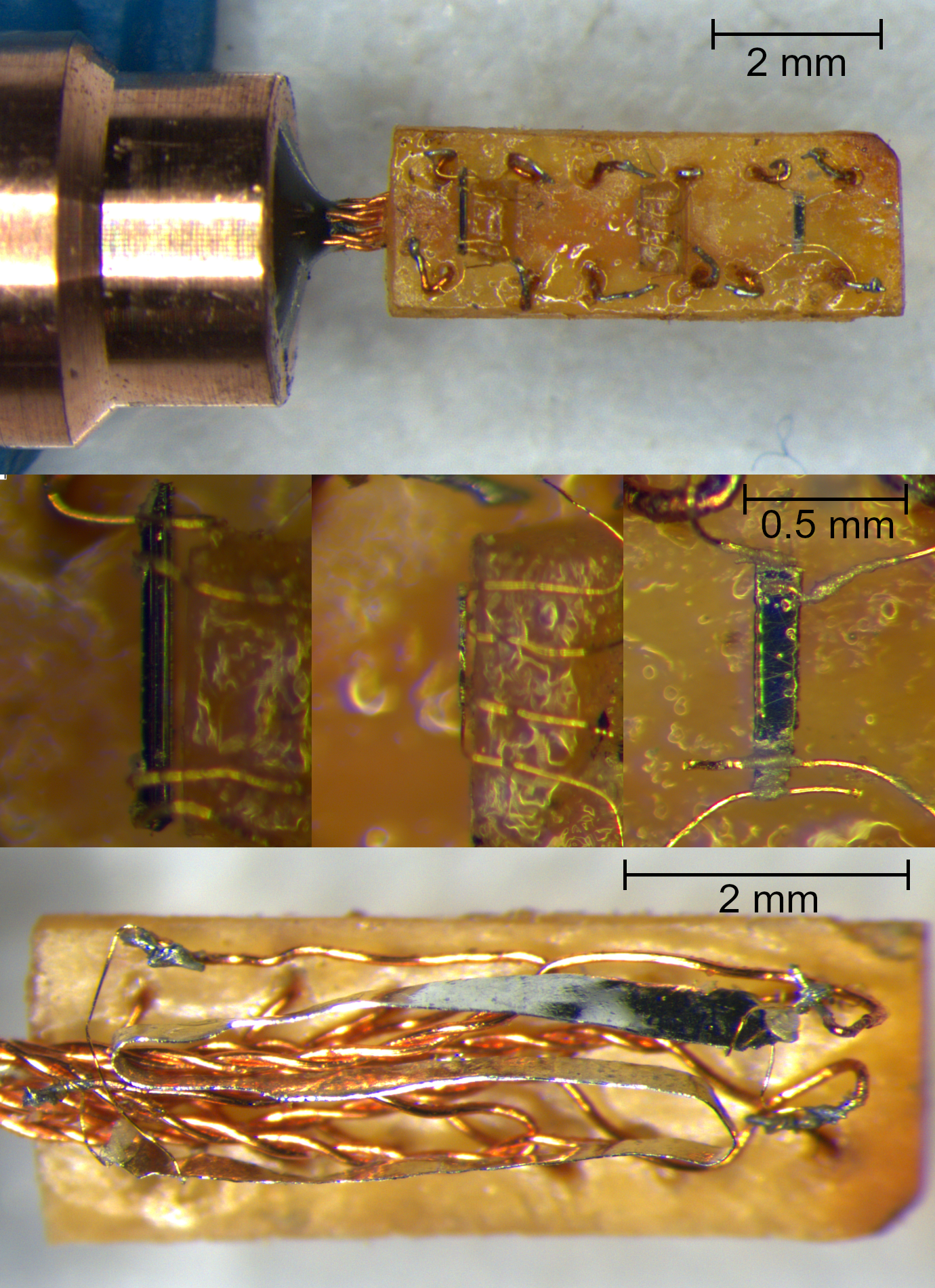}
	    \caption{Piston-cylinder cell feed-through with the samples mounted. Top: overview of the entire sample-carrying platform. Middle: individual samples. Samples with the $c$ axis along the field direction were supported by rectangular blocks. Bottom: the other side od the platform with a strip of tin (Sn) serving as a manometer.}
        \label{sfig:samples}
\end{figure}

In the main text we show resistivity data obtained using the same batch of \CRA\ single crystals that was used in Refs.~\cite{khim2021,hafner2022,landaeta2022,kibune2022,ogata2023}. These have the resistivity ratio of 2.0 between 300\,K and 0.5\,K, and their $C/T$ value extrapolates to $\gamma=$1.2\,J\,K$^{-2}$\,mol$^{-1}$ at zero temperature. In section V of this Supplemental Material, in Fig.~\ref{sfig:T0_new_samples}, we show resistivity for a higher quality batch, also used in Ref.~\cite{semeniuk2023} (resistivity ratio of 2.6, $\gamma=$0.7\,J\,K$^{-2}$\,mol$^{-1}$). Figure~\ref{fig:T0_data}(a) displays heat capacity for an even newer batch (resistivity ratio of 2.7, $\gamma=$0.2\,J\,K$^{-2}$\,mol$^{-1}$). We prepared the samples for resistivity measurements by mechanically cutting and polishing crystals into elongated rectangular bars, preserving naturally formed edges and facets when possible (the typical dimensions can be seen in Fig.~\ref{sfig:samples}). Electrical contacts with gold leads were made by spot-welding and were additionally reinforced by silver-loaded epoxy (DuPont 6838). Separate measurements on \CRA\ samples of well-defined dimensions gave resistivity values at 300\,K of $\rho=125\,\micro\Omega$\,cm for the original batch. In this work, we used this room-temperature resistivity value as a reference when converting the measured voltage readings into the absolute resistivity. A conversion using the estimated geometric factors of the actual samples used in this work gave consistent resistivity values, but with large uncertainties due to the small sizes and geometric irregularities of the samples.

\subsection{Hydrostatic pressure}
\begin{figure}[t]
        \includegraphics[width=\linewidth]{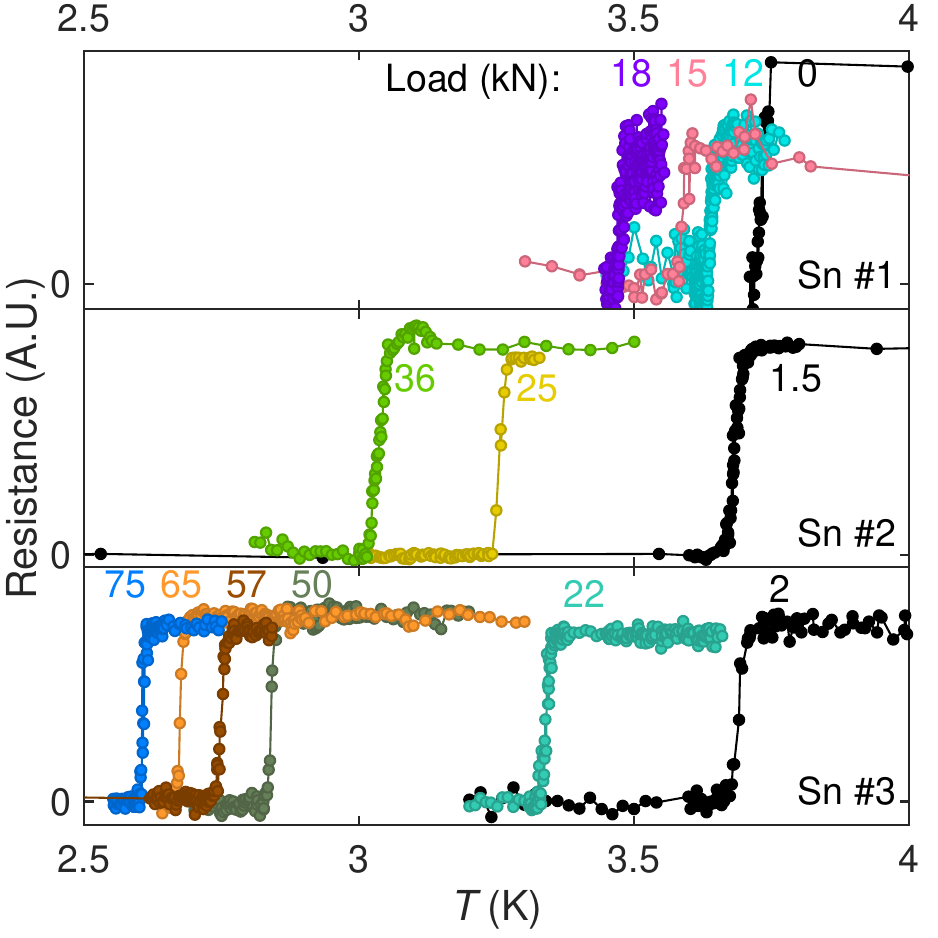}
	    \caption{Superconducting transitions of tin (Sn) manometers at different pressure cell load values. The three panels represent the different Sn samples used throughout the experiment, labelled as Sn~\#1, Sn~\#2, and Sn~\#3. We only show the data for the loads at which we conducted the low-temperature measurements on \CRA.}
        \label{sfig:Sn}
\end{figure}

\begin{figure}[t]
        \includegraphics[width=\linewidth]{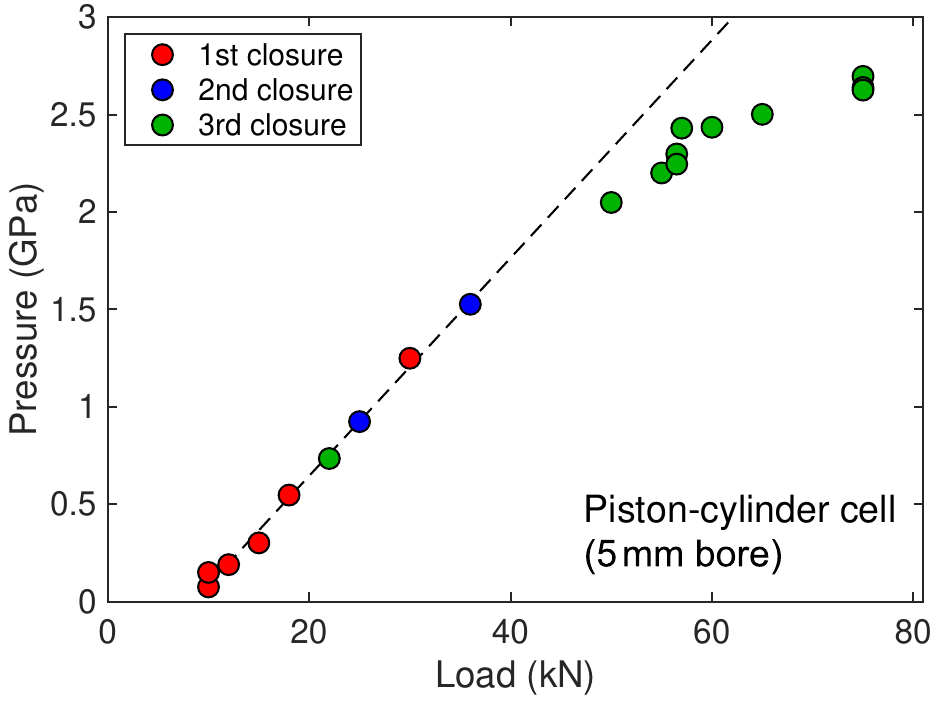}
	    \caption{Pressure inside the piston-cylinder cell as a function of applied load. The low-temperature properties of \CRA\ were not measured at every point shown. The linear regime up to $\sim50$\,kN is emphasized with the dashed line.}
        \label{sfig:load_v_p}
\end{figure}

High-pressure conditions were achieved with a piston-cylinder cell produced by C\&T Factory Co. Ltd. The cell featured a hybrid design with a a beryllium-copper jacket and non-magnetic nickel-chrome-aluminium insert with a 5.0\,mm diameter bore. Three samples of \CRA\ were placed inside the pressure cell, along with a strip of Sn acting as as a manometer (see Fig.~\ref{sfig:samples}). Pressure was determined from the superconducting (SC) transition temperature of the Sn sample, according to the calibration in Ref.~\cite{smith1969}, with careful prior elimination of remnant magnetic field of the PPMS cryostat. The SC transitions of different Sn manometers used throughout the experiment are shown in Fig.~\ref{sfig:Sn}. The pressure-load curve is shown in Fig.~\ref{sfig:load_v_p}. Polyethylsiloxane (PES1) silicone oil was used as a pressure transmitting medium.

\subsection{Uncertainty in pressure}

For each pressure point, the superconducting transition temperatures of the Sn manometer were obtained via resistivity measurements in PPMS cryostats. However, thermal cycling of the cell can cause a slight change of pressure, adding to the uncertainty in pressure for a subsequent measurement in a dilution refrigerator. At 2.67 and 2.27\,GPa, we measured pressure before and after the measurements in a dilution refrigerator and found a decrease of, respectively, 0.06 and 0.05\,GPa. Taking into account the uncertainty of $\pm$0.02--0.03\,GPa based on the width of the SC transition of Sn, we estimate the overall uncertainty in pressure to be $\pm$0.05\,GPa.

\subsection{Resistivity measurements at low temperatures and high magnetic fields}

Resistivity measurements at temperatures below 2\,K were conducted in dilution refrigerators via the four-point AC technique, using lock-in amplifiers. We used low-temperature transformers (medium-frequency cryogenic transformers supplied by CMR-direct) with the gain of 100 for the noise-free signal amplification. For the runs at 2.50, 2.67, and 2.23\,GPa, low-noise pre-amplifiers were additionally used at room temperature for signal filtering. The excitation current was in the 0.5--1\,$\micro$A range, across which the variation of \Tc\ was below 1\,mK, indicating no significant Joule heating. The temperature of the experiment was monitored using two thermometers, located at the mixing chamber of the dilution refrigerator (in a field-compensated region) and directly at the pressure cell. The pressure cell was coupled to the mixing chamber with a securely clamped bundle of silver wires of the combined cross section area of approximately 1.5\,mm$^{2}$.

At temperatures below 20\,mK, the heat capacity of beryllium-copper increases dramatically~\cite{karaki1997}. Applying magnetic field shifts this increase to higher temperatures. This effect resulted in the thermal lag becoming more pronounced at lower temperatures and higher magnetic fields, which was counteracted by conducting temperature sweeps in both directions and at a reasonably low rate, typically between 200 and 50\,mK/h, when measuring below 1\,K.

\begin{figure}[t]
        \includegraphics[width=\linewidth]{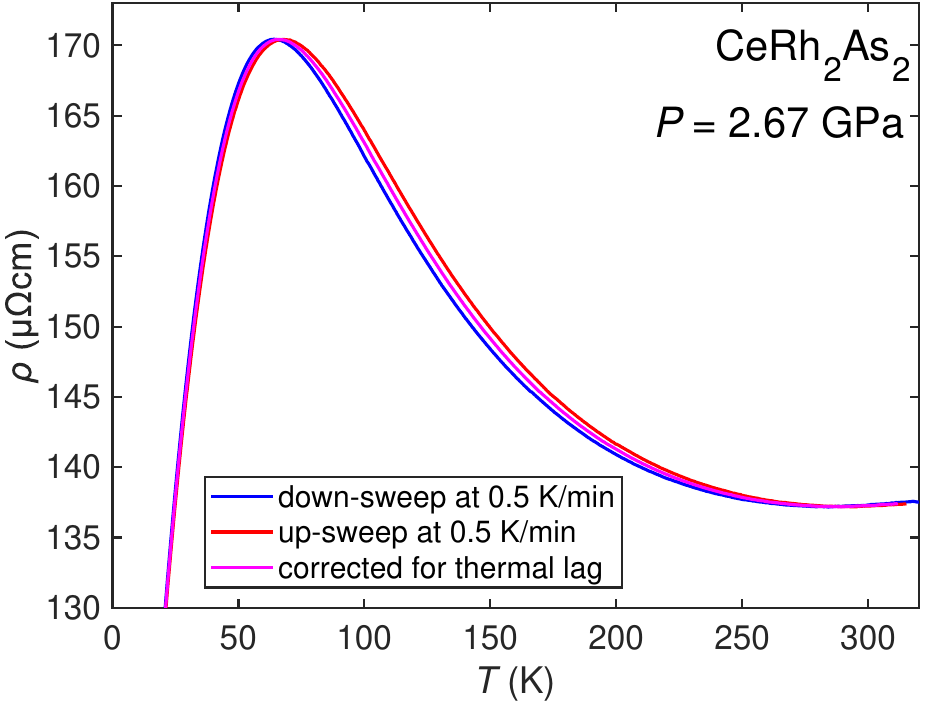}
	    \caption{Downward and upward temperature sweeps of resistivity of \CRA\ at high pressure showing the difference due to the thermal lag of the pressure cell. The two sweeps were combined in order to obtain the corrected set of data.}
        \label{sfig:tlag}
\end{figure}

Measurements at higher temperatures (between 2 and 300\,K) were conducted with Quantum Design PPMS cryostats using the AC transport option with a 0.1\,mA excitation current. Due to a significant heat capacity of the pressure cell at temperatures above 50\,K, conducting temperature sweeps at a 0.5\,K/min rate (the lowest rate deemed practical) still resulted in a thermal lag of up to 4\,K. The extent of the thermal lag is illustrated in Fig.~\ref{sfig:tlag}. The data plotted in Fig.~2(a) were processed to eliminate the thermal lag, which was determined from a combination of upward and downward temperature sweeps as also shown in Fig.~\ref{sfig:tlag}.

We defined the SC critical temperature \Tc\ as the point of maximum slope in the resistivity drop. Defining \Tc\ as the 50\% threshold of the resistivity drop did not change the results in any meaningful way. We used the point of maximum curvature of $\rho(T)$ as the criterion of \To. The derivatives of $\rho(T)$ were determined using the Savitzky-Golay filter with a second order polynomial.

For the plots of $\rho(T)$ in Fig.~2(b) and Fig.~3(b), the data were smoothed via binning with the bin sizes of 1 and 5\,mK respectively. In Fig.~4(a), the same 1 and 5\,mK binning was used for the 0\,T and 6\,T data, respectively.

\subsection{Heat capacity measurements under pressure}

The heat capacity measurements were carried out in a copper-beryllium piston-cylinder cell with a 2\,mm bore diameter~\cite{nicklas2015a}. Four samples of \CRA\ of the combined mass of 23.56\,mg and a piece of lead (Pb), serving as the pressure gauge, were sealed in a Teflon cap inside the bore. Flourinert-72 was used as the pressure-transmitting medium. The heat capacity of the whole pressure cell with the samples was determined by a compensated heat-pulse technique in a dilution refrigerator. To obtain the absolute heat capacity of the \CRA\ samples, a previously measured addenda heat capacity of the entire setup, including the pressure cell but without the \CRA\ samples, was subtracted. The value of pressure inside the cell was obtained from the shift of the SC transition temperature of the lead piece inside the cell compared to a piece outside the cell. The superconducting transition temperature of lead was determined by magnetization measurements using an MPMS (Quantum Design).

\section{Experimental protocol}

The pressure cell for resistivity measurments was initially loaded with three samples of \CRA\ which we label as Sample~1, Sample~2, Sample~3. Sample~1 and Sample~2 are not relevant in the scope of this manuscript. Sample~3 is the source of the resistivity data in Fig.~\ref{fig:ZeroFieldData}, Fig.~\ref{fig:T0_data}, and Fig.~\ref{fig:normal_state_summary}. It was mounted with the crystallographic $c$ axis perpendicular to the axis of the cell, which was also the direction of the applied magnetic field. The initial alignment was confirmed to be well within 1\degree\ using an optical microscope.

The chronological sequence of applied pressures (in GPa) was the following: 0.30, 0.19, 0.55, 0.92, 1.53, 0.73, 2.05, 2.50, 2.67, 2.27. The cell was opened after measurements at 0.30, 0.55, and 1.53\,GPa in order to fix problems with Sample~1. After 1.53\,GPa, Sample~1 was replaced with Sample~4, from the batch used in Ref.~\cite{semeniuk2023}. Sample~4 is the source of the data in Fig.~\ref{sfig:T0_new_samples}, displayed in section V below. Each time after the cell was opened and closed again, a small load of up to 2\,kN was first applied in order to seal the cell without applying any significant pressure, and a measurement was conducted to ensure that the ambient pressure results are reproduced.

For each new pressure point we conducted two kinds of measurements. A measurement in a PPMS cryostat was typically done first, in order to determine the pressure inside the cell from the SC transition temperature of the Sn manometer (shown in Fig.~\ref{sfig:Sn} and to probe the zero-field resistivity of \CRA\ in the 2\,K--300\,K temperature range. Next, a measurement at lower temperatures and with applied magnetic fields were conducted in a dilution refrigeration cryostat.

\section{On the different values of \Tc\ in resistivity and heat capacity}

There is an apparent difference in \Tc\ between resistivity data in Fig.~2(b) and heat capacity data in Fig.~3(a), with the heat capacity reporting a lower \Tc\ than resistivity, even though a higher quality sample was used for the former measurement. Such a discrepancy is not uncommon in unconventional superconductors, where \Tc\ is generally very sensitive to various deviations from the perfect crystallinity. Crystalline defects and impurities lead to a spatially non-uniform distribution of \Tc\ across a sample, and its values can even be enhanced above the nominal \Tc\ due to a favourable local strain, for instance. When probing resistivity, the reading drops to zero as soon as there is a superconducting path between the voltage electrodes, which may happen even if only a minor fraction of the sample is superconducting. The superconducting anomaly seen in heat capacity measurements always reflects the entire distribution of \Tc\ in a sample.

\section{Field-temperature phase diagram at ambient pressure}

\begin{figure}[b]
        \includegraphics[width=\linewidth]{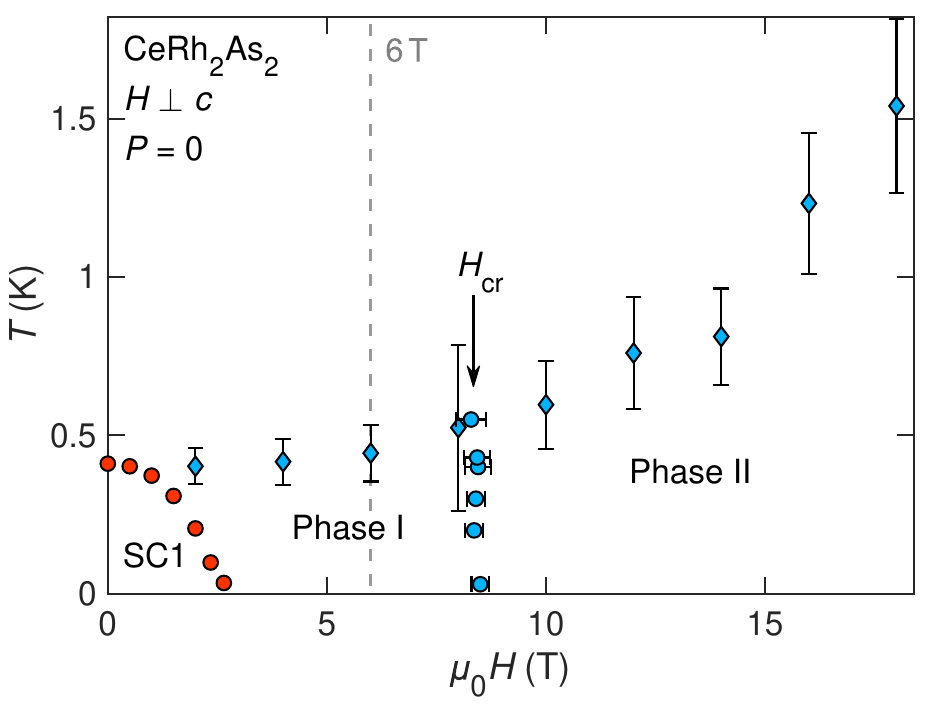}
	    \caption{Temperature-in-plane-field phase diagram of \CRA\ at ambient pressure for the sample used in this work (Sample 3).}
        \label{sfig:Phase_diagram_ambient}
\end{figure}

In Fig.~\ref{sfig:Phase_diagram_ambient} we show the ambient-pressure phase diagram of \CRA\ in the space of temperature and in-plane magnetic field, obtained for the sample measured by resistivity in the present work (Sample~3). The phase diagram is in a good agreement with the previously published ones~\cite{hafner2022,mishra2022}. Namely, phase I extends until $\sim8.5$\,T and $T_{0}(H_{\perp})$ has a weakly positive slope for low fields.

\section{Additional data supporting the suppression of \To with pressure}

\begin{figure}[h]
        \includegraphics[width=\linewidth]{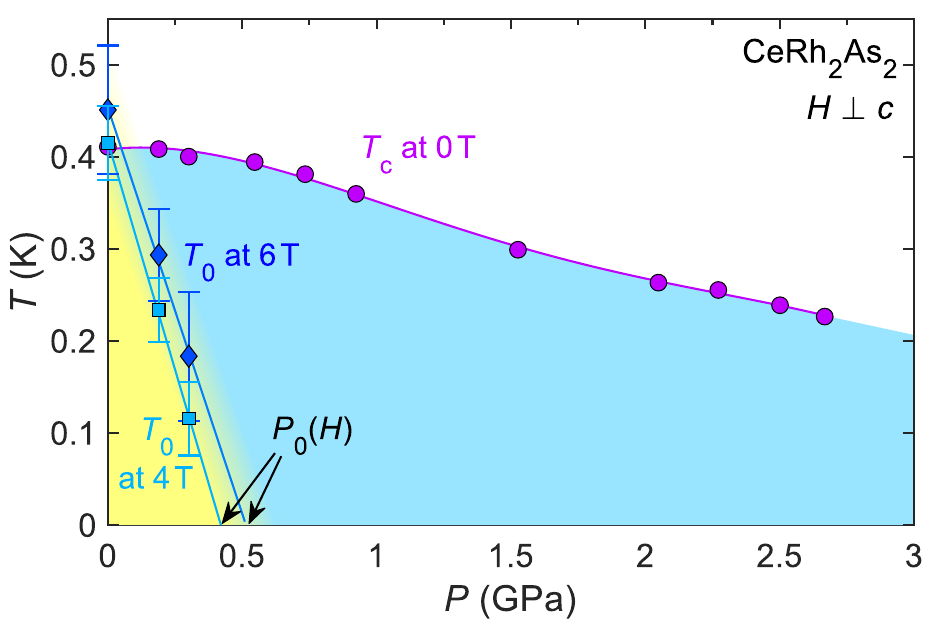}
	    \caption{Combined pressure-temperature phase diagram showing the behaviors of \Tc\ at zero field and of \To\ at 4 and 6\,T.}
        \label{sfig:PD_with4T}
\end{figure}

In Fig.~\ref{sfig:PD_with4T} we extend the data shown in Fig.~\ref{fig:normal_state_summary} of the main text by including the pressure dependence of \To\ at a 4\,T in-plane magnetic field, obtained from resistivity measurement. The critical pressure of Phase~I at this field is 0.41\,GPa, rather than 0.5\,GPa at 6\,T, which is in line with \To\ being slightly smaller at 4\,T than at 6\,T at ambient pressure (see Fig.~\ref{sfig:Phase_diagram_ambient}). 

\begin{figure}[h]
        \includegraphics[width=\linewidth]{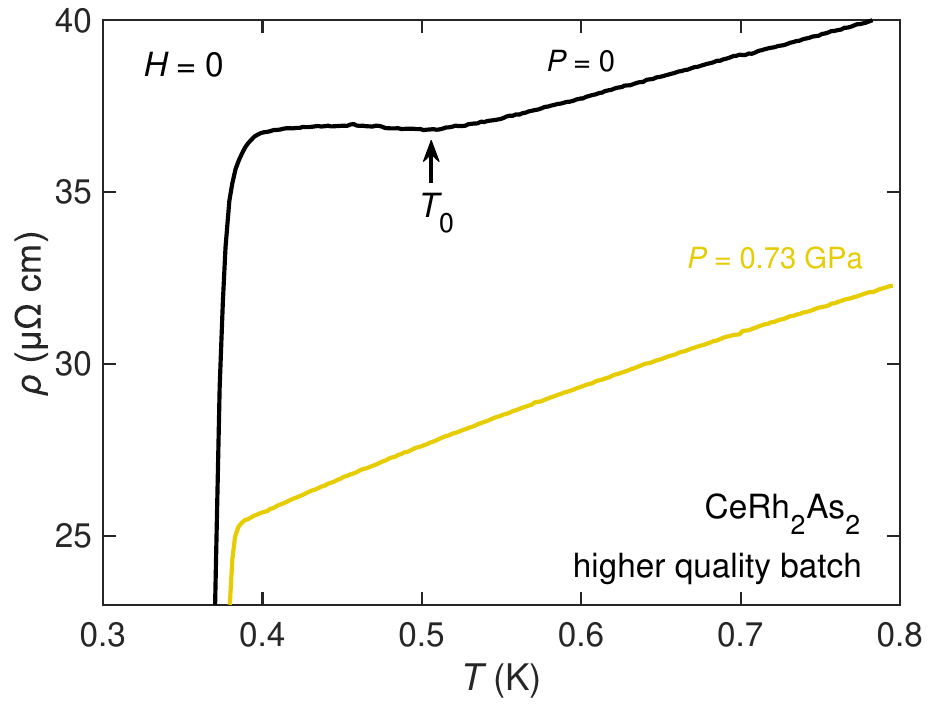}
	    \caption{Temperature dependence of resistivity $\rho(T)$ of improved \CRA\ crystals at ambient pressure and 0.73\,GPa. The reference value of $\rho=95\,\micro\Omega$\,cm at 300\,K was used for obtaining the absolute resistivity values.}
        \label{sfig:T0_new_samples}
\end{figure}

Higher quality crystals of \CRA~\cite{semeniuk2023} became available during the study. A sample from that batch (Sample~4) showed clearly resolved resistive signatures of \To\ and \Tc\ at zero field. We only have a limited set of high-pressure data for this new sample, but they show (Fig.~\ref{sfig:T0_new_samples}) that the resistivity upturn at \To\, clearly visible at ambient pressure, is fully gone at 0.73\,GPa. Furthermore, \Tc\ is enhanced at 0.73\,GPa compared to the ambient pressure value.

\section{Power-law analysis of the temperature dependence of resistivity}

In this section, we provide a more detailed explanation of the power-law analysis of low-temperature resistivity of \CRA. In conventional Fermi-liquid metals, electron-electron interactions dominate charge carrier scattering at low temperatures, leading to a standard temperature dependence of the resistivity of the form $\rho(T)=\rho_{0}+AT^2$, where $\rho_{0}$ is the residual resistivity and $A$ is a temperature-independent coefficient, acting as a measure of the strength of electron-electron interactions~\cite{jacko2009}. In non-Fermi-liquid metals, the $T^2$ term changes to a more general $T^n$ term, where the exponent $n$ (now different from 2) depends on the scattering processes involved. In order to keep dimensions consistent, the power law function can be rewritten as $\rho(T)=\rho_{0}+A^*(T/T_{\mathrm{ref}})^n$, where $T_{\mathrm{ref}}$ is some characteristic temperature. The $A$ coefficient is changed into $A^*$, which now has units of resistivity. One can easily notice that $A^*$ is equal to the increment of resistivity between $T=0$ and $T_\mathrm{ref}$, irrespective of $n$, and hence the coefficient still correlates with the effective quasiparticle mass. Therefore, as a QCP is approached, besides the exponent $n$ being distinctly different from 2, the $A^*$ coefficient is expected to increase dramatically. In our analysis, we use the pressure dependence of $A^*$ and $n$ to probe the proximity of the system to a QCP.

At low temperatures, $\rho(T)$ of \CRA\ cannot be well-approximated by a fixed power-law exponent over a broad temperature range of the order of 1\,K, which is why we consider a smaller temperature interval of 0.25\,K, which still allows a reliable fitting. To be most sensitive to the QCP-related effects, we also have to examine the lowest possible temperature region, unless the \To\ transition is present, in which case, the next best solution is to consider an interval just above the onset of \To, which we took to be 0.15\,K above the maximum curvature point. In the case of 0.55\,GPa, when only the onset of \To\ is visible, we took 0.15\,K as the lower bound for the fitting interval. For the higher pressures, the fitting interval started at the lowest achieved temperature. The temperature intervals used for each analyzed pressure point are listed in Table~\ref{stab:PL_fit_intervals} and the resulting parameters are presented in Fig.~\ref{fig:normal_state_summary}(b) of the main text.

\begin{table}[h]
\centering
\begin{tabular}{ |c|c|c| } 
 \hline
 $P$ (GPa) & $T_{1}$ (K) & $T_{2}$ (K) \\
 \hline
 0 & 0.601 & 0.851 \\ 
 0.19 & 0.443 & 0.693 \\ 
 0.30 & 0.333 & 0.583 \\ 
 0.55 & 0.150 & 0.400 \\ 
 0.73 & 0.035 & 0.285 \\ 
 0.92 & 0.053 & 0.303 \\ 
 1.52 & 0.091 & 0.341 \\ 
 2.05 & 0.086 & 0.336 \\ 
 2.50 & 0.031 & 0.281 \\ 
 2.67 & 0.030 & 0.280 \\ 
 \hline
\end{tabular}
\caption{Temperature intervals for the power law fitting at different pressures. $T_{1}$ and $T_{2}$ are the lower and upper bounds of the intervals.}
\label{stab:PL_fit_intervals}
\end{table}

\begin{figure}[h]
        \includegraphics[width=\linewidth]{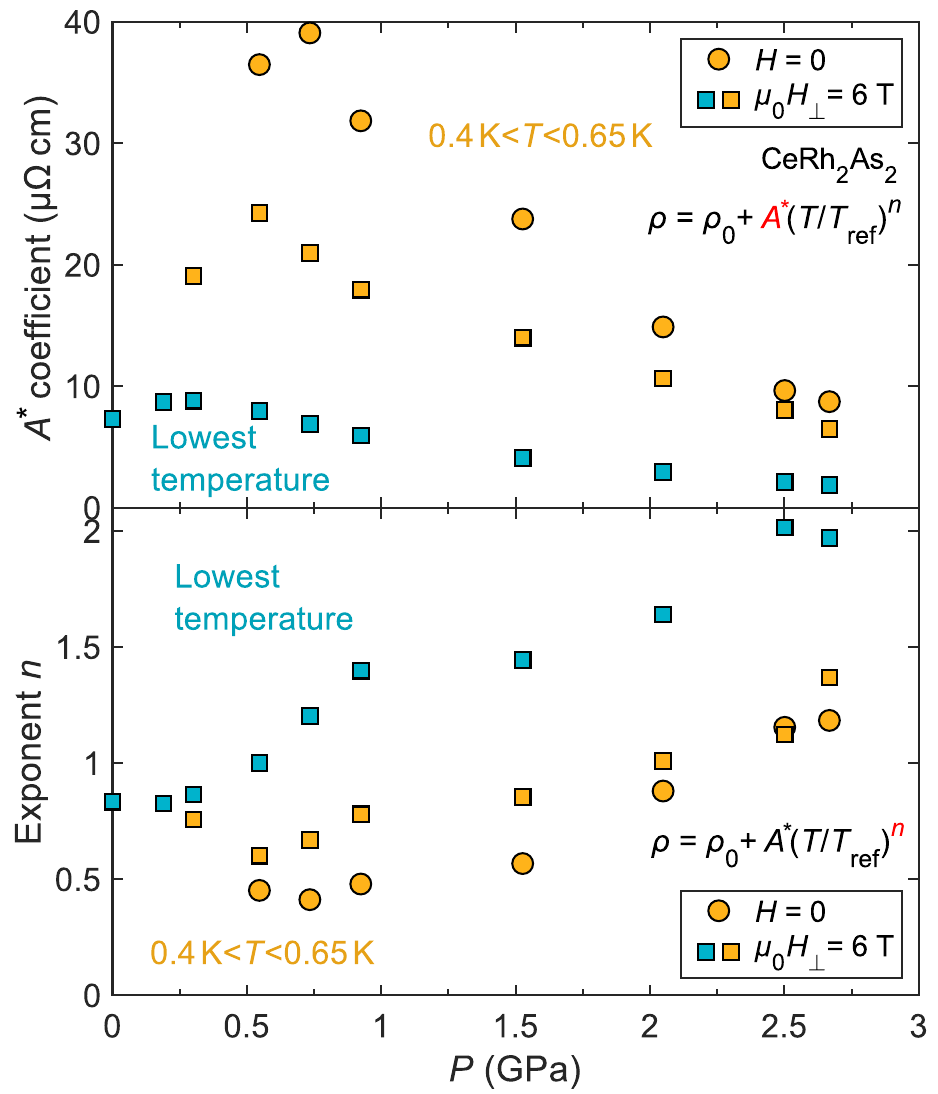}
	    \caption{Pressure dependence of the $\rho(T)$ power law coefficients for the different fitting conditions. We used a 0.25\,K-wide fitting interval with $T_{\mathrm{ref}}=0.3$\,K for the lowest accessible temperature range (see Table~\ref{stab:PL_fit_intervals}) and $T_{\mathrm{ref}}=0.65$\,K for the fixed interval between 0.4\,K and 0.65\,K.}
        \label{sfig:PL_fits}
\end{figure}

To check whether the results are robust, we conducted the same analysis for a fixed higher-temperature interval $0.4\,\mathrm{K}<T<0.65\,\mathrm{K}$ (with $T_{\mathrm{ref}}=0.65$\,K), for both 6\,T and zero-field data (only for pressures where neither \Tc\ nor \To\ influence the data in the specified temperature range). The outcome is shown in Fig.~\ref{sfig:PL_fits} along with the results for lower temperatures. We found that the same trend of increasing $n$ and strongly decreasing $A^*$ also holds at these temperatures, both for 0 and 6\,T, validating the low-temperature result.

\bibliography{CeRh2As2_pressure_normal}
\end{document}